\documentclass[hidelinks, 11pt]{article}

\usepackage[left=1in, right=2in, top=1in, bottom=1in, includeheadfoot, nohead]{geometry}
\usepackage[right,mathlines,running]{lineno}
\usepackage{caption}
\usepackage{titling}
\usepackage{authblk}
\usepackage{libertine}
\usepackage{libertinust1math}
\usepackage[T1]{fontenc}
\makeatletter
\renewcommand{\maketitle}{\bgroup\setlength{\parindent}{0pt}
\begin{flushleft}
{\large\@title}
\vspace{0.2cm}\\
\@author\\
\end{flushleft}\egroup}
\makeatother
\usepackage{fancyhdr}
\usepackage{lastpage}
\usepackage{booktabs}

\usepackage{hyperref}
\usepackage[usenames,dvipsnames,svgnames,table]{xcolor}
\usepackage{soul}
\newcommand{\hrefc}[3][blue]{\href{#2}{\color{#1}{#3}}}
\definecolor{myLink}{HTML}{08519c}
\definecolor{mySig}{HTML}{737373}

\definecolor{myBlue}{HTML}{deebf7}

\usepackage{amsthm}
\usepackage{thmtools}
\usepackage[unq]{unique}
\usepackage{wrapfig}
\usepackage{float}
\declaretheoremstyle[
spaceabove=0pt, spacebelow=0pt,
headfont=\normalfont\bfseries,
notefont=\mdseries, notebraces={(}{)},
bodyfont=\normalfont,
postheadspace=0em,
qed=
]{mystyle}
\declaretheorem[style=mystyle, numbered=no, shaded={bgcolor=myBlue, textwidth=0.5\textwidth}]{Significance Statement}

\usepackage[bottom]{footmisc}
\usepackage{sectsty}
\definecolor{mySect}{HTML}{969696}
\sectionfont{\large\sc\color{mySect}}
\subsectionfont{\normalsize\sc\color{mySect}}

\usepackage{amsmath}
\usepackage{bm}
\usepackage{mathtools}
\usepackage{graphicx}
\usepackage{capt-of}
\usepackage[export]{adjustbox}
\usepackage{tabularx}

\usepackage[style=nature]{biblatex}
\title{\sc \Large Interspecific dispersal constraints suppress pattern formation in metacommunities}

\author[1]{\hrefc[myLink]{https://orcid.org/0009-0002-4216-2174}{\small Patrick Lawton}}
\author[2,3]{\hrefc[myLink]{https://orcid.org/0000-0001-9138-3593}{\small Ashkaan K. Fahimipour}}
\author[4,*]{\hrefc[myLink]{https://orcid.org/0000-0003-0773-3779}{\small Kurt E. Anderson}}

\affil[1]{Biophysics Graduate Program, University of California, Riverside, CA, USA}
\affil[2]{Department of Biological Sciences, Florida Atlantic University, Boca Raton, FL, USA}
\affil[3]{Center for Complex Systems and Brain Sciences, Florida Atlantic University, Boca Raton, FL, USA}
\affil[4]{Department of Evolution, Ecology, \& Organismal Biology, University of California, Riverside, CA, USA}

\begin{document}
\vspace*{0.5in}
\maketitle
\noindent$^*$Correspondence to:
\texttt{\hrefc[myLink]{mailto://kurt.anderson@ucr.edu}
{\footnotesize kurt.anderson@ucr.edu}}\\
\vspace{0.25cm} {\color{mySect}\hrule} \vspace{0.25cm}
\sloppy
\noindent
\textcolor{mySig}{
Decisions to disperse from a habitat stand out among organismal behaviors as pivotal drivers of ecosystem dynamics across scales.
Encounters with other species are an important component of adaptive decision-making in dispersal, resulting in widespread behaviors like tracking resources or avoiding consumers in space.
Despite this, metacommunity models often treat dispersal as a function of intraspecific density alone.
We show, focusing initially on three-species network motifs, that interspecific dispersal rules generally drive a transition in metacommunities from homogeneous steady states to self-organized heterogeneous spatial patterns.
However, when ecologically realistic constraints reflecting adaptive behaviors are imposed --- prey tracking and predator avoidance --- a pronounced homogenizing effect emerges where spatial pattern formation is suppressed.
We demonstrate this effect for each motif by computing master stability functions that separate the contributions of local and spatial interactions to pattern formation. We extend this result to species rich food webs using a random matrix approach, where we find that eventually webs become large enough to override the homogenizing effect of adaptive dispersal behaviors, leading once again to predominately pattern forming dynamics.
Our results emphasize the critical role of interspecific dispersal rules in shaping spatial patterns across landscapes, highlighting the need to incorporate adaptive behavioral constraints in efforts to link local species interactions and metacommunity structure.
}


\vspace{0.25cm} {\color{mySect}\hrule} \vspace{0.5cm}
\section*{Introduction}
Organismal behavior plays a pivotal role in shaping ecosystems across scales. Dispersal in particular exerts a profound influence, affecting everything from the availability of local resources \cite{stier2014larval, fahimipour2015colonisation, gil2017social} and the distribution of predators \cite{hein2011predators} to the structure of ecological networks \cite{pillai2011metacommunity, fahimipour2014dynamics}, and the dynamics of metacommunities \cite{amarasekare2008spatial, gross2020modern, anderson2021body, fahimipour2022sharp}. However, understanding the effects of dispersal on metacommunity dynamics has been made difficult by the variation and complexity of behaviors that ultimately influence whether an organism leaves a given habitat \cite{hein2020algorithmic, gross2020modern, anderson2021body}. Often, these dispersal decisions are influenced by encounters with other species, including with resources, competitors, or consumers \cite{kreuzinger2022population, legrand2015ranking, fronhofer2018bottom, fahimipour2023wild}. Despite the observable dependencies of dispersal on local diversity and species interactions, many metacommunity models still treat dispersal as function of intraspecific density alone \cite{amarasekare2008spatial, abrams2007habitat, amarasekare2010effect, anderson2021body}.

The consideration of dispersal rates that respond to species interactions is needed as previous studies suggest they may have substantial impacts on metacommunity dynamics \cite{fronhofer2018bottom, abrams2007habitat, amarasekare2010effect, quevreux2021predator, fahimipour2022sharp, brechtel2019far, pedersen2016nonhierarchical}. These dispersal responses are often represented as ``cross-diffusion’’ terms, where the movement of one species is influenced by the spatial gradient of another species' density or abundance \cite{turchin1998quantitative, murray2003mathematical, cantrell2004spatial}. A recurring finding in recent spatial models incorporating cross-diffusion is the heightened sensitivity of spatially homogeneous steady states to spontaneous pattern formation driven by dispersal, i.e. Turing instabilities \cite{vanag2008crossdiffusion, zemskov2013turing, fanelli2013turing, madzvamuse2017crossdiffusion, krause2021modern}. To ensure that these models capture ecologically relevant phenomena, it is important to incorporate cross-diffusion terms that reflect the signs of interspecific interactions, such that prey would avoid predators or consumers would actively track prey \cite{vanag2008crossdiffusion}. However, when such constraints on cross-diffusivity have been implemented in ecological models, they have mostly been applied to two-species systems \cite{tsyganov2004solition, fahimipour2022sharp}, limiting the range of potential dynamical behaviors \cite{krause2021modern}. These simplified models fall short in capturing the dynamics of larger species interaction networks, leaving critical gaps in our understanding of the link between interaction-driven dispersal and the self-organized community structures that emerge at realistic scales \cite{hata2014dispersal, brechtel2019far, hayes2023persistence, pillai2011metacommunity, gouhier2010synchrony}.

\begin{figure*}[t!] 
\includegraphics[width=0.95\textwidth]{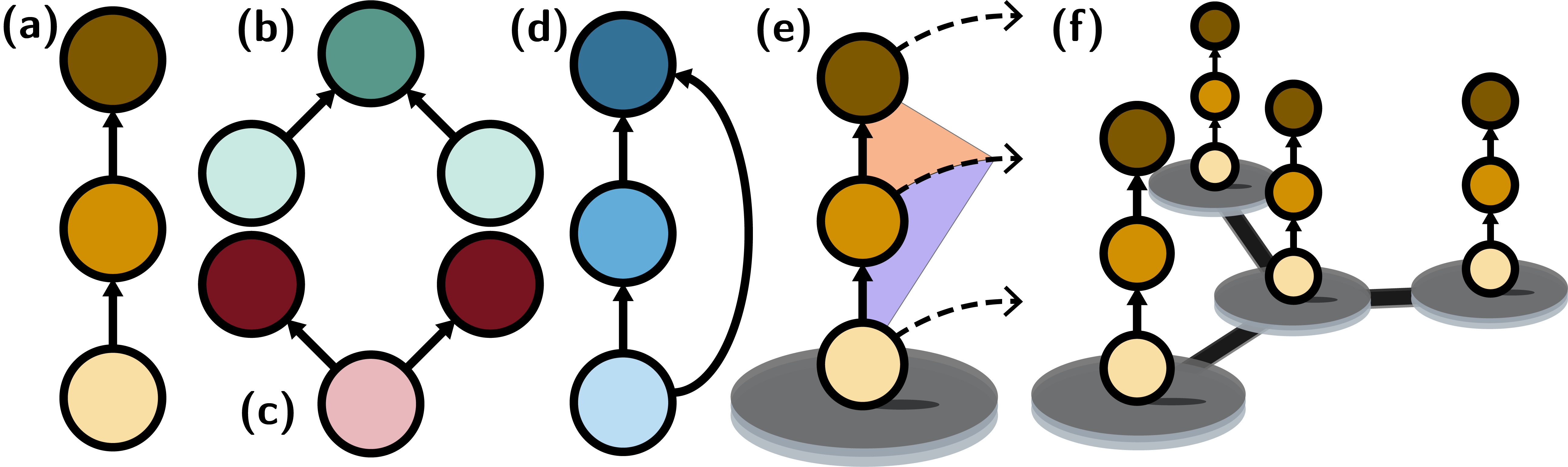}
\caption{Food web motifs and metacommunity connectivity}{
\footnotesize
\textcolor{mySig}{
Solid arrows denote feeding relationships. Light, medium and dark shading represents primary producers, intermediate consumers, and top consumers, respectively. Motifs are labeled as (a) food chain, (b) apparent competition, (c) resource competition, (d) intraguild predation. N > 3 food webs are not shown, but may include one or more of these motifs. (e) Species may disperse from patches (dashed lines) in response to other species' densities (shaded areas). For webs (a)-(c), a maximum of $n_{\text{cross}}=4$ such interspecific dispersal responses are possible. When constrained, dispersal responses are opposite in sign from feeding relationships such that dispersal is higher for a species with locally high densities of its predators and low densities of its prey. (f) Dispersal connects habitats in spatial networks to create metacommunities. 
}}
\label{fig:diagram}
\end{figure*}
Here, we assess the susceptibility of model metacommunities to spatial pattern formation, both with and without ecologically-relevant adaptive constraints applied to the sign of cross-diffusion terms (i.e., predator avoidance and prey tracking).
Our results consistently indicate that increasing the prevalence of unconstrained cross-diffusion facilitates spatial pattern formation. 
However, the introduction of ecological constraints that regulate cross-diffusion diminishes or even eliminates the tendency for pattern formation. 
We first examine ecologically relevant three-species interaction motifs (Fig.~1a-d) via a rigorous computational scan of both non-spatial (trophic) and spatial (dispersal) model parameters.
We then extend our analysis to systems comprising more than three species using a more efficient random matrix approach, showing that findings in our three species motifs qualitatively hold when extended to larger metacommunities. 

\section*{modeling and analysis framework}
We employ a metacommunity modeling framework in order to study the effects of cross-diffusion on spatial food web dynamics. We assume these dynamics are governed by a set of deterministic reaction-diffusion equations on a spatial network describing species interactions and dispersal in separable terms. Nodes in the spatial network represent habitat patches where species interactions determine both local food web dynamics and decisions to emigrate to neighboring patches. The dynamics of species $i$ on a given patch $k$ are then captured by
\begin{align} \label{eq:metacommunity}
    \dot{x}_i^k = f_i(\bm{x}^k) - \sum_l{\text{L}_{kl}D_i({\bm{x}^l})},
\end{align}
where the generally non-linear function of species $i$'s biomass density $f_i(\bm{x}^k)$ and $D_i({\bm{x}^k})$ define the rates of trophic interactions and dispersal, respectively, and are assumed to be identical across all habitat patches. Connections between patches are represented by the Laplacian matrix $\text{L}_{kl}=\delta_{kl}\sum_l\text{A}_{kl} - \text{A}_{kl}$, with $\text{\bf A}$ the network's adjacency matrix. As exact solutions to Eq 1. rarely exist, either direct simulation or analysis of the linearized system can be employed to understand the long term dynamics. 

We consider the dynamics of Eq. 1 following small perturbations $\delta \bm{x}^k$ from a spatially homogeneous, nontrivial steady state $\bm{x}^\ast$. These perturbations can be decomposed over eigenmodes of $\bf{L}$, each with a corresponding Laplacian eigenvalue $\kappa$ analogous to the wavenumber in continuous space \cite{nakao2010turing, fahimipour2022sharp, hata2014dispersal, brechtel2018master}. 
The allowed spatial signatures of all eigenmodes, as encoded in the respective eigenvectors, is predetermined by the structure of $\bf{L}$, unlike in continuous space where perturbations may be decomposed over arbitrarily high wavenumbers.
To evaluate the potential dynamical behavior across arbitrary patch networks, we therefore refrain from specifying a particular form of $\bf{L}$ and instead consider $\kappa$ as a real valued, positive parameter of arbitrarily high magnitude.
The exponential growth rate $\lambda$ of the $n$th mode for each species is then given by eigenvalues of 
\begin{align}
    \bf{J} = \bf{P} - \kappa \bf{C},
\end{align}
where the local Jacobian
\begin{align} \label{eq:Pij}
    \text{P}_{ij} = \frac{\partial f_i(\bm{x}^k)}{\partial x_j^k} \bigg\rvert_{\bm{x}^\ast}
\end{align}
encodes the linearized non-spatial component of the model (i.e., trophic interactions) while the connectivity matrix
\begin{align} \label{eq:C}
    \text{C}_{ij} = \frac{\partial D_i(\bm{x}^k)}{\partial x_j^k} \bigg\rvert_{\bm{x}^\ast}
\end{align}
encodes the linearized spatial response of species $i$ to species $j$ near $\bm{x}^\ast$ \cite{brechtel2018master, fahimipour2022sharp}.

The long term behavior of Eq 1. can be qualitatively understood via the distribution of all possible eigenmodes' maximum growth rates 
\begin{align}
    \lambda(\kappa) = {\rm Ev}_{\max}\left({\bf J}\right),
\end{align}
where the right hand side denotes the leading eigenvalue of the matrix ${\bf J}$, variably referred to as the dispersion relation \cite{murray2003mathematical, scholes2019a} or the master stability function \cite{fahimipour2022sharp, brechtel2018master}. 
We delineate possible metacommunity outcomes into three qualitatively different dynamical behaviors based on the form of $\lambda(\kappa)$.
The first, which we label 'stable' dynamics (\textit{st}), occur when perturbations decay on an isolated patch (i.e. $\lambda(0)<0$) and the homogeneous state $\bf{x}^\ast$ is similarly maintained on any spatial patch network \cite{fahimipour2022sharp}, such that $\lambda(\kappa)<0$ for all $\kappa>0$.
In contrast, 'unstable' dynamics (\textit{us}) are characterized by an initially positive master stability function (i.e. $\lambda(0) > 0$) such that stable coexistence of all species beginning from $\bm{x}^\ast$ is impossible in isolation or on any spatial network.
Finally, 'pattern forming' dynamics (\textit{pf}) occur when $\lambda(0) < 0$ but crosses $0$ at some critical $\kappa$ value. In these systems, certain spatial networks will cause the system to self-organize to a heterogeneous state where species exhibit variation in densities --- either static or oscillatory --- across patches in the spatial network. 

To assess the tendency of ecological systems to exhibit any of the particular dynamical behaviors outlined above, it is important to account for the effects of parameter variation in Eq. 1. We quantify this tendency as the robustness of a given behavior to variation in the parameter space of Eq. 1 \cite{li2017incoherent, scholes2019a}. Separating the effects of trophic and dispersal interactions, we define the local robustness $\omega(\bf{P})$, spatial robustness $\omega(\bf{C})$, and total robustness $\omega(\bf{P},\bf{C})$ as fractions of the appropriate parameter spaces yielding a particular behavior (\textsl{see Supplement}). This definition is conceptually akin to that of feasibility domains discussed elsewhere in ecological literature \cite{grilli2015feasibility}. Local robustness thus quantifies the fraction of possible interaction models, encoded in $\bf{P}$, which result in a particular dynamical outcome under a set of behavioral assumptions on dispersal, encoded in $\bf{C}$. Similarly, spatial robustness quantifies the fraction of considered dispersal behaviors with a common outcome under a set of assumptions on trophic interactions. Finally, the total robustness quantifies the frequency of a given outcome as both local and spatial parameters vary. 
 
To construct the connectivity matrix $\bf{C}$, we first assume diagonal entries are strictly positive to accommodate mass-action diffusive effects. Off-diagonal elements $\text{C}_{ij}$ may appear for each pairwise food web interaction $\text{P}_{ij} \neq 0$ resulting in a total of $n_{\text{cross}}$ interspecific dispersal responses. To impose ecologically reasonable (i.e., adaptive) constraints we then restrict the signs of these off diagonals as 
\begin{align} \label{eq:constraint}
    \text{sgn}(\text{C}_{ij})=-\text{sgn}(\text{P}_{ij}), ~i \neq j,
\end{align} 
such that the effects of prey tracking and predator avoidance are accounted for \cite{vanag2008crossdiffusion, tsyganov2004solition}. Rather than assume a specific form for $D_i({\bm{x}^k})$, we focus on the linearized spatial responses in Eq.~4. While this prevents us from performing direct simulation of the system, it allows for a comprehensive numerical sampling scheme to approximate the robustness of dynamical behaviors, for which analytical predictions become intractable for $N > 2$ species \cite{krause2021modern}.

\section*{3 species motifs}
We start by considering the effects of interspecific dispersal rules on conventional models of 3-species ecological interaction motifs (Fig.~1). Local population dynamics on patch $k$ for each $N = 3$ interaction motif are defined by a generalized system of equations:

\begin{align} \label{eq:3spectrophic}
f_i(\bm{x}^k) = & x_i \Bigl[
    \smash{\underbrace{
    r_i \left( 1 - \frac{x_i}{K_i} \right)
    }_{\substack{\text{primary} \\ \text{production}}}}
    ~-~\smash{\underbrace{
    \sum_{j \neq i}{\frac{A_{ij} x_j}{B_{ij} + x_i}}
    }_{\substack{\text{loss from} \\ \text{predation}}}}
    ~+~\smash{\underbrace{
    \sum_{j \neq i}{\frac{A_{ji} e_{ji} x_j}{B_{ji} + x_j}}
    }_{\substack{\text{gain from} \\ \text{predation}}}}
    ~- \smash{\underbrace{
    {d_ix_i}
    }_{\text{mortality}}}
    \Bigr] \\ \nonumber & \\ \nonumber &
\end{align}
where $r$ and $K$ are the intrinsic population growth rate and carry capacity, respectively, of producers while $A$, $B$, $e$ and $d$ are the attack rate, half-saturation constant, conversion efficiency, and mortality coefficient, respectively, of consumers. 
The relevant gain and loss terms due to predation interactions appear according the interaction motif. This corresponds to ecological models with logistic producer growth, and consumers that exhibit type-II functional responses and density-dependent mortality.

\begin{figure*}[t!]
\includegraphics[width=0.95\textwidth]{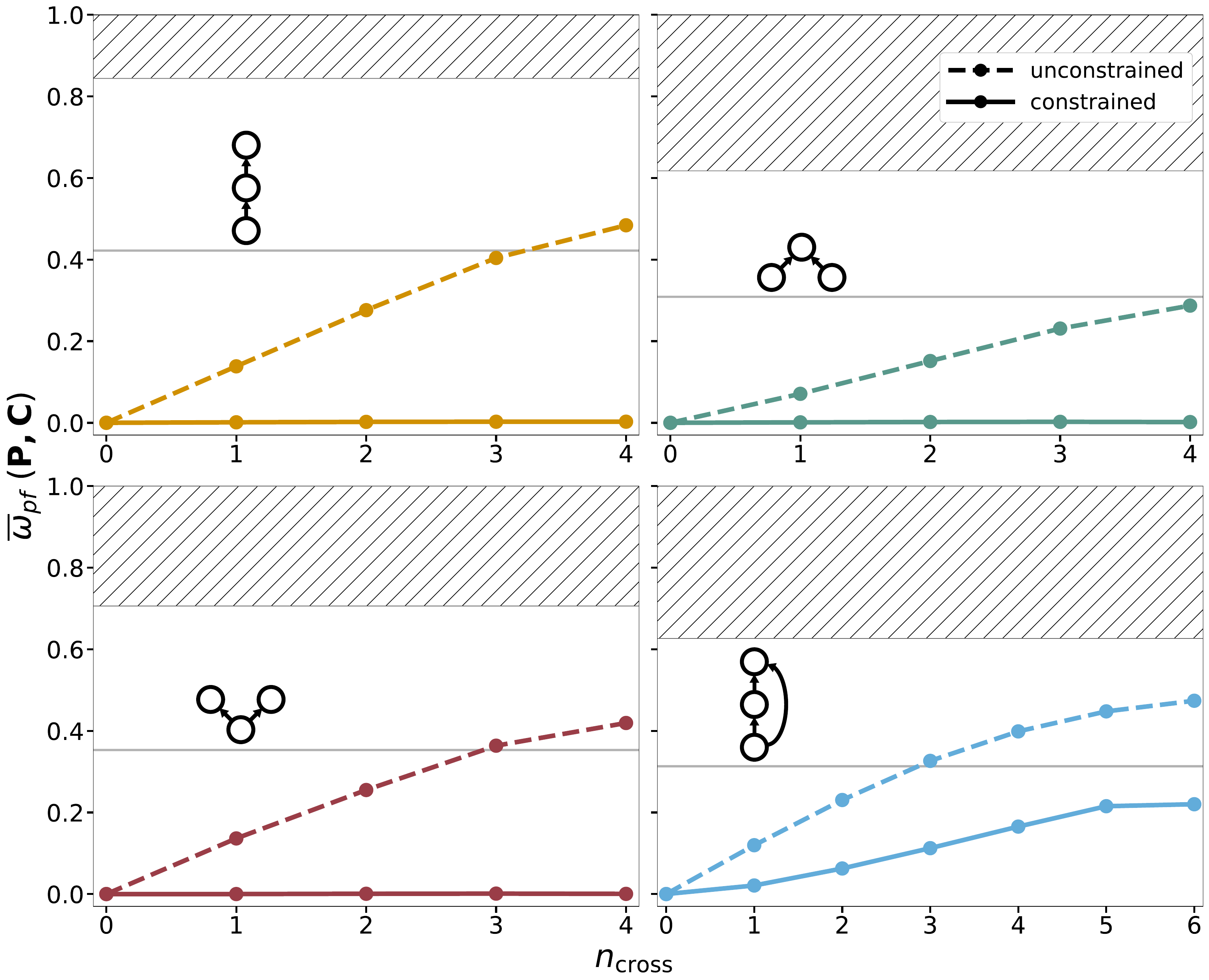}
\caption{Robustness of pattern formation in food web motifs}{
\footnotesize
\textcolor{mySig}{
Total robustness of pattern forming dynamics $\omega_{pf}(\bf{P},\bf{C})$ shown is averaged over permutations of $n_{\text{cross}}$ interspecific dispersal responses for a given local interaction motif. The robustness of unstable dynamics depends exclusively on local interactions and is indicated by the hatched area, providing an upper bound on $\omega_{pf}(\bf{P},\bf{C})$. The robustness of stable dynamics (not shown) is simply $1-(\omega_{pf}+\omega_{us})$. Thin horizontal line indicates the threshold above which pattern formation becomes the most robust dynamical behavior. 
}}
\label{fig:3spectotalrobust}
\end{figure*}

In accordance with previous studies on 2 species systems, for all considered 3 species systems we find that the inclusion of interspecific spatial responses has the potential to induce pattern formation which is not possible with intraspecific dispersal alone (i.e. $n_{\text{cross}}=0$). This potential increases monotonically with the prevalence of cross diffusion (i.e. with $n_{\text{cross}}$) such that pattern forming dynamics eventually become the most robust to variation in both local and spatial interaction parameters for 3 out of 4 food web motifs (Fig.~2). That is, a random parameterization of Eq. \ref{eq:3spectrophic} is highly likely to yield a steady state $\bf{x}^\ast$ susceptible to pattern formation if spatial interactions between species are common and no restrictions exist on the relative weight or sign of these interactions. The increase in $\omega_{pf}(\bf{P},\bf{C})$ with such unconstrained spatial responses is qualitatively similar across all interaction motifs, despite consisting of markedly different trophic structures and propensities for locally unstable dynamics. 

However, under the behavioral constraints on dispersal in Eq. \ref{eq:constraint}, the robustness of pattern forming dynamics is significantly diminished. Regardless of the local interaction motif or the value of $n_{\text{cross}}$, stable dynamics become more robust to parameter variation than pattern forming dynamics. For systems with only 2 directly interacting species, the constrained value of $\omega_{pf}(\bf{P},\bf{C})$ falls to nearly zero. This indicates that a locally stable steady state of food webs a-c in Fig. \ref{fig:diagram} will be maintained regardless of the underlying spatial network or the particular model parameters, so long as the behaviors of prey tracking and predator avoidance are strictly adhered to. For the case of 3 direct interactions in food web D, pattern forming dynamics remain likely for a randomly parameterized metacommunity, but significantly less so relative to the case of unconstrained cross-diffusion.  Thus, Fig. \ref{fig:3spectotalrobust} reflects an increased tendency for the system in Eq. \ref{eq:3spectrophic} to maintain spatial homogeneity rather than transition to a heterogeneous state when spatial interactions follow reasonable expectations for adaptive behaviors.

While the cumulative effect of adding interspecific dispersal terms on metacommunity behavior is equivalently given by the first moment, i.e. the mean, of either robustness metric, higher moments can shed light on differential impacts for either spatial or local robustness. Thus, we compute the second moment, the variance, and the third moment, the skewness, about the shared mean for the distributions of local and spatial robustness at each value of $n_{\text{cross}}$. To draw a direct comparison of these quantities between each robustness metric, we consider the distributions over parameter samples of Eq. \ref{eq:3spectrophic} which yield food webs that are both feasible and stable in isolation, i.e. we exclude webs which yield unstable dynamics. 

\begin{figure*}[t!]
\includegraphics[width=0.95\textwidth]{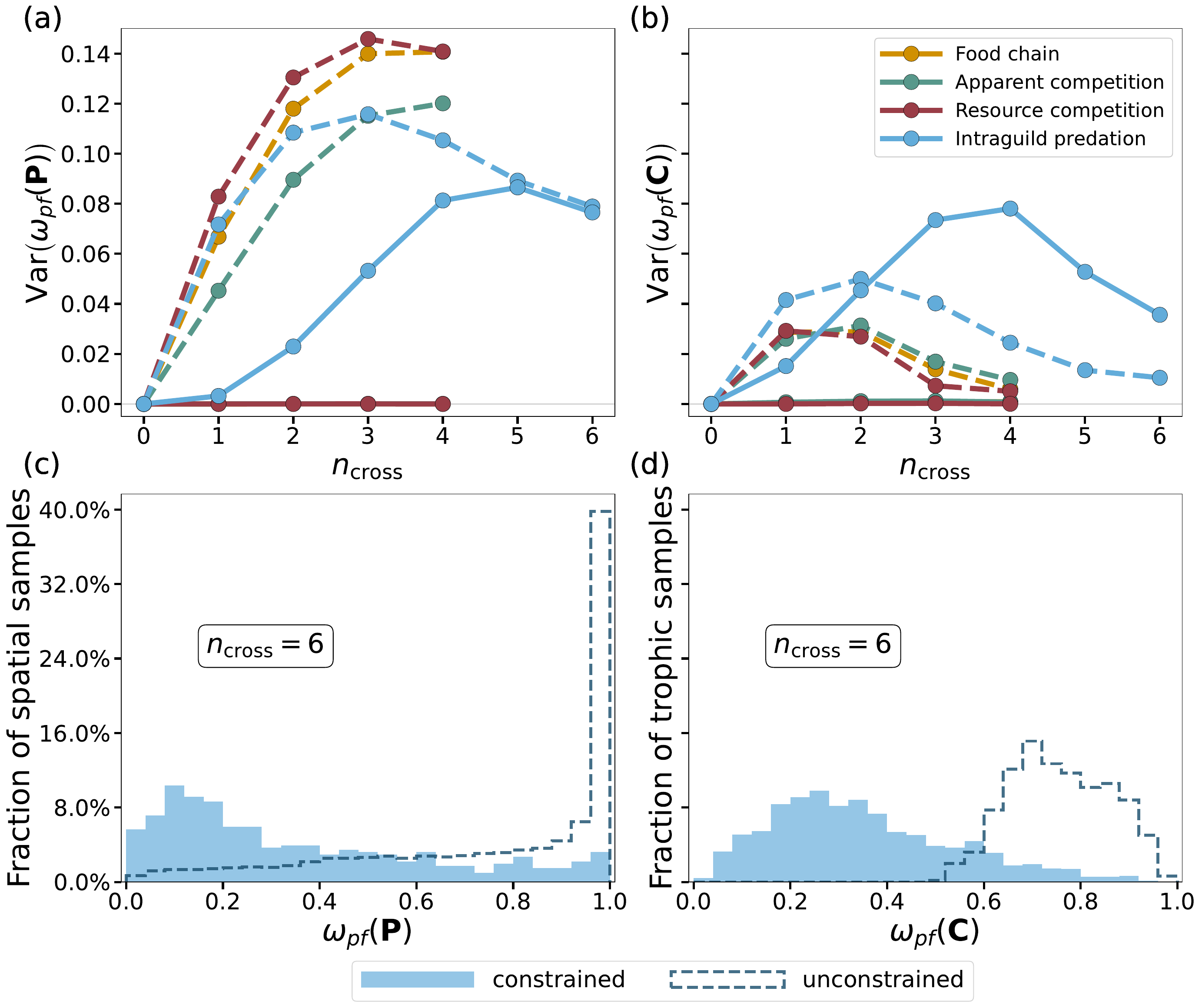}
\caption{Distributions of spatial and local robustness}{
\footnotesize
\textcolor{mySig}{
Local (a, c) and spatial (b, d) robustness distributions considered exclude locally unstable model parameterizations, such that $\omega_{pf}$ may reach $1$ as opposed to Fig. \ref{fig:3spectotalrobust}. Without constraints on interspecific dispersal (dotted line), the sensitivity of pattern forming behavior to dispersal rates (a) is significantly higher than the sensitivity to local interaction parameters (b). Dispersal constraints (solid line) render these sensitivities comparable while diminishing or even prohibiting pattern formation. Example distributions (c) and (d) for the intraguild predation motif show that, at high $n_{\text{cross}}$, stabilization of homogeneous equilibria via dispersal constraints is reflected most prominently in the local robustness distribution.
}}
\label{fig:robuststats}
\end{figure*}

Beginning with the second moment, the variance provides a heuristic measure of model sensitivity to variation in species' spatial responses when computed for $\omega(\bf{P})$ or sensitivity to variation in the local parameters of Eq. \ref{eq:3spectrophic} when computed for $\omega(\bf{C})$. As seen in Fig. \ref{fig:robuststats}a, the robustness of pattern forming dynamics is highly dependent on the species' spatial responses, encoded in the near-equilibrium dispersal rates of Eq. \ref{eq:C}. For all local interaction motifs, this sensitivity is maximized at an intermediate value of $n_{\text{cross}}$ approximately where the difference in total robustness between stable and pattern forming dynamics is minimized (Fig. \ref{fig:3spectotalrobust}). For higher values of $n_{\text{cross}}$, susceptibility to pattern formation becomes increasingly common for initially stable steady states of Eq. \ref{eq:3spectrophic}, and thus $\text{Var}(\omega_{pf}(\bf{P}))$ decreases as the dynamical behavior becomes less dependent on the particular set of dispersal parameters. In comparison, the sensitivity to local parameters of Eq. \ref{eq:3spectrophic} shown in Fig. \ref{fig:robuststats}b is relatively low, and the maxima of $\text{Var}(\omega_{pf}(\bf{C}))$ occur prior to that of the $\omega_{pf}(\bf{P})$ distribution. This indicates that dispersal is the more decisive factor in determining dynamical behavior relative to local interaction conditions for systems where species respond to one another spatially without any constraints on such responses.

The imposition of the constraints in Eq. \ref{eq:constraint} significantly alters these sensitivities. For motifs with only two direct interactions, the variance of either distribution is approximately zero as stable dynamics become virtually guaranteed. In contrast, the intraguild predation motif (Fig.~1D) maintains a high sensitivity to near-equilibrium dispersal rates while $\text{Var}(\omega_{pf}(\bf{C}))$ increases such that the local and spatial sensitivities become comparable. Thus, when spatial responses are possible between all species comprising our $N=3$ metacommunity, the dynamical outcome of Eq. \ref{eq:metacommunity} is highly sensitive to the choice of parameters and susceptibility to pattern formation remains common for an initially stable food web under cross diffusive constraints, even while stable dynamics become generally preferred.

Turning to the third central moment, the skewness indicates the tendency for randomly selected dispersal rates to yield values of local robustness higher or lower than the total robustness (i.e. the mean) when computed for $\omega(\bf{P})$, and similarly indicates this tendency for randomly selected local interaction parameters when computed for $\omega(\bf{C})$. We find that for low values of $n_{\text{cross}}$, the robustness of pattern forming dynamics tends to be lower than the average shown in Fig. \ref{fig:3spectotalrobust} in the absence of cross-diffusive constraints from Eq. \ref{eq:constraint}. Imposition of the constraints exacerbates this effect, in particular with respect to the choice of local interaction parameters, and at higher $n_{\text{cross}}$ the skew towards relatively low $\omega_{pf}$ is comparable for both robustness distributions. In the absence of these constraints, however, the local robustness distribution skews towards relatively high $\omega_{pf}$, while the spatial robustness becomes symmetric about the mean. 

Overall, when dispersal responses between species are few, ecologically motivated dispersal constraints result in a more significant shift towards preference of stable dynamics in the $\omega(\bf{C})$ distribution compared to the $\omega(\bf{P})$ distribution. However, when interspecific dispersal responses are prevalent, this stabilization of homogeneous equilibria is instead reflected most prominently in the local robustness distribution. This highlights the strong potential of interspecific dispersal responses, and their particular dependence on food web interactions, to determine the dynamical behavior of spatially explicit metacommunities.

\section*{Large random metacommunities}
To understand the impacts of interspecific dispersal rules on the dynamics of large metacommunities, we employed a method to generate random Jacobian matrices based on species interactions defined by the niche model \cite{williams2000simple}. To construct the network Jacobians, we first generate a food web topology by drawing niche values randomly from a uniform distribution for each of $N$ species in the metacommunity. These values depict each species' position along a 1-dimensional trophic niche axis. Niche ranges are then determined for each species by randomly sampling values from a Beta distribution with parameters that depended on the desired connectance \cite{williams2000simple}. For species pair $(i, j)$, if the niche value of $j$ falls within the range of $i$, then $i$ is designated the consumer and $j$ as a resource. The corresponding Jacobian $\mathbf{P}$ entries are modified accordingly: $\text{P}_{ij}$ receives a positive entry drawn from a folded normal distribution $\sim \left| \mathcal{N}(0, \sigma) \right|$, while $\text{P}_{ji}$ receives a value drawn from $\sim -\left| \mathcal{N}(0, \sigma) \right|$. Finally, the diagonal entries $\text{P}_{ii}$ are set to $-1$, reflecting density-dependent effects and self-regulation. Resulting webs were examined to ensure that paths exist between all species in the food web (i.e., there are no disconnected sub-webs), otherwise the web was discarded and re-generated.
To populate the connectivity matrix $\bf{C}$, we define an auxiliary parameter $q$ which specifies the probability that the effect of species $j$ on $i$ will lead to a uniformly distributed dispersal kernel, such that $\text{C}_{ij} = \text{Bern}(q)~\cdot~\sim \mathcal{U}(-1, 1)$, with the sign of $\text{C}_{ij}$ optionally constrained by Eq. \ref{eq:constraint}.

The effects of cross-diffusion observed in our 3 species model with fully-specified trophic interactions qualitatively hold when extended to larger metacommunities with interactions strengths in Eq. \ref{eq:Pij} randomly assigned. Firstly, the potential for pattern formation increases dramatically with the prevalence of interspecific dispersal responses $q$. In the absence of constraints on these responses, locally stable $\bf{P}$ matrices are guaranteed to be susceptible to pattern formation at sufficiently high $q$ for most food web sizes $N$. A stronger tendency towards pattern formation is observed relative to our 3 species results, reflected in a higher maximum robustness and in a lower number of interspecific responses needed to reach said maximum. Secondly, imposing the ecological constraints of Eq. \ref{eq:constraint} significantly diminishes $\omega_{pf}(\bf{P},\bf{C})$, even eliminating the possibility of pattern formation for metacommunities with relatively few species. While the robustness of pattern forming dynamics still scales with $q$ for large $N$, pattern forming dynamics never become guaranteed as it does for unconstrained cross diffusion. Thus, the increased propensity for a homogeneous state under strict behavioral constraints on interspecific dispersal is a generic feature of our results, regardless of the number of species.

\begin{figure*}[t!]
\includegraphics[width=0.95\textwidth]{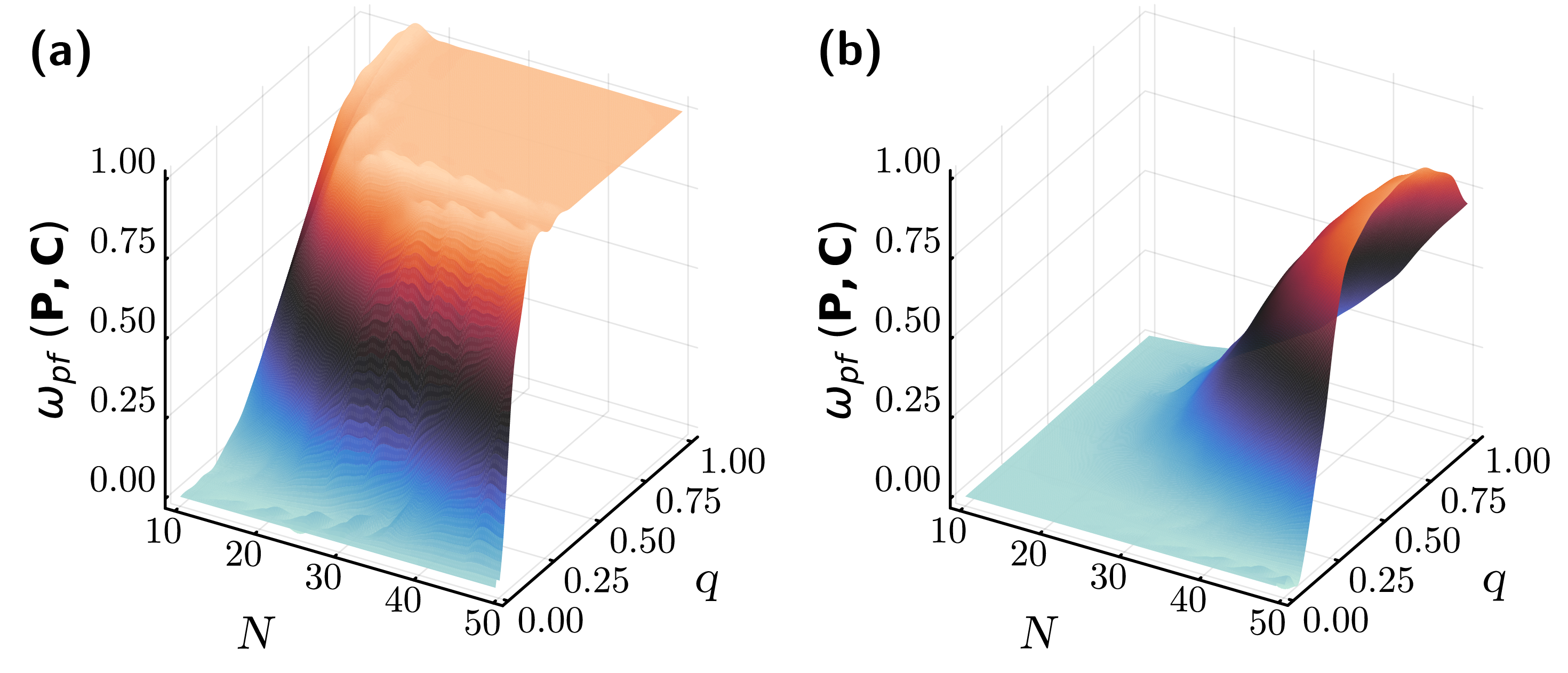}
\caption{Pattern formation in large metacommunities}{
\footnotesize
\textcolor{mySig}{
As in Fig.\ref{fig:robuststats}, locally unstable model parameterizations are excluded, such that $\omega_{pf}$ may reach $1$. Vertical axis and color indicate total robustness of pattern forming behavior in randomly generated metacommunities (methods), with $q$ the probability of a food web interaction having a corresponding interspecific dispersal response either without (a) or with (b) the ecologically motivated constraints in Eq. \ref{eq:constraint}.
}}
\end{figure*}

\section*{Discussion}

In this paper we show that dispersal driven by interspecific interactions has profound effects on spatial-pattern formation tendencies in metacommunities. 
Ecological studies commonly simplify dispersal as a linear function of intraspecific density with similar rates for all species, assumptions which tend to stabilize spatially uniform equilibrium states \cite{murray2003mathematical, cantrell2004spatial}.
Generally, we find that assuming dispersal rates which respond locally to both intra- and inter- specific densities significantly increases susceptibility to pattern formation. 
However, when interspecific dispersal responses are constrained to follow empirically observed adaptive behaviors \cite{kreuzinger2022population, legrand2015ranking, fronhofer2018bottom, fahimipour2023wild}, i.e. prey tracking and predator avoidance, pattern formation is dramatically suppressed. 
This effect is observed regardless of food web size, although for sufficiently large number of species pattern forming dynamics take precedent over homogeneous dynamics with or without adaptive dispersal constraints applied (Fig.~4).

The three-species interaction motifs investigated here showed minimal pattern formation when interspecific dispersal was constrained. The intraguild predation module was a notable exception as it exhibited significant pattern formation in both the unconstrained and constrained scenarios (Fig.~2). This motif is distinctive among the three-species motifs we examined because it involves interactions links spanning multiple trophic levels, potentially introducing additional dispersal feedbacks that can destabilize the system. Interestingly, we observed limited pattern formation in larger metacommunities with fewer than  approximately 20 species (Fig.~4B). While all motifs could potentially co-occur as building blocks of larger metacommunities, it appears that the homogeneous dynamics observed in certain motifs appear to dominate less speciose metacommunities. The presence of pattern-forming omnivory in such cases might be stabilized by other motifs or factors \cite{anderson2021body} or only become a frequent module in larger metacommunities. Studies that construct large food webs around specific interaction motifs could elucidate which ones are most important for pattern formation and represents an important future research direction \cite{gellner2012reconciling, gellner2016consistent}.

The spatially heterogeneous metacommunity dynamics observed in our models may take different forms. Heterogeneous spatial patterns in metacommunities may include static differences in species densities --- Turing-like patterns --- and localized oscillations that may be synchronous or asynchronous across patches \cite{murray2003mathematical, hata2014dispersal, hata2014sufficient, pedersen2016nonhierarchical}. Certain forms of spatially asynchronous variability promoting species persistence \cite{abbott2011dispersal, gouhier2010synchrony, hayes2023persistence, anderson2018effects, guill2021self}. Internally generated patterns have been observed in nature \cite{rietkerk2008regular}, although empirical links between pattern formation and ecosystem resilience are few \cite{zhang2023self, maron1997spatial}. This is particularly the case of large scale metacommunity dynamics \cite{guill2021self}, where variation may only manifest across longer temporal and spatial scales.

Regardless of the exact form, pattern formation mechanisms generate variation among patches even without underlying variation in the abiotic environment. However, environmental variation within metacommunities is commonplace and underpins notable metacommunity paradigms \cite{leibold2004metacommunity}. Models of large random metacommunities show that spatial environmental variation can promote community persistence, inverting the classic relationship between community complexity and instability \cite{anderson2021body, gravel2016stability, mougi2016food}. Environmental variation may also drive dispersal, generating another path for altering cross-diffusion terms. By increasing the number and/or strength of community linkages, cross-diffusion terms would likely increase complexity and therefore reduce pattern formation when spatial environmental variation is present. Furthermore, high dispersal among certain species can provide sufficient dispersal to increase stability when overall dispersal is low \cite{anderson2021body,brechtel2019far}; this effect could be promoted by adaptive dispersal behavior \cite{amarasekare2007spatial, mougi2019adaptive}. Whether constrained or unconstrained cross-diffusion leads to greater metacommunity complexity or influences pattern formation in spatially heterogeneous habitats is an open question.  

Our models assume that dispersal decisions are based on local resource or predator densities. Emigrants dispersing from a patch then ``resettle'' in neighboring patches without an assessment of the new conditions there. In contrast, fitness-dependent models implement dispersal as a function of differences between both the "donor" and "recipient" habitats (e.g., \cite{amarasekare2007spatial, amarasekare2010effect, abrams2007habitat}. These alternative representations of dispersal lead to different outcomes regarding species coexistence and distributions among habitats \cite{gross2020modern, gramlich2016influence}. Which model of dispersal best approximates the behavior of real organisms remains an open question, and likely depends on the focal system in question (reviewed in \cite{gross2020modern}). However, it is probably unlikely that individuals can rapidly assess differences between community conditions at the scale over which most metacommunities operate (but see \cite{fronhofer2017information, ponchon2022informed}).    

Still, our assumption that dispersal is triggered directly by local resource or predator densities may not be general to all ecological systems. While we restrict interactions to only exist between consumers and resources, competing species are also known to drive each other’s emigration decisions \cite{fronhofer2015condition}. Species responding to the presence of competitors may be especially common under contest competition (e.g. competition for space \cite{calcagno2006coexistence}). Given that direct dispersal responses to competitors would add cross-diffusion terms similar to ``unconstrained'' ones in our model, we expect the inclusion of such responses to further increase the propensity of pattern forming dynamics. 

Furthermore, species may use both intra- and interspecific social cues when dispersing. Congregation behaviors may be reflected in directional biases in movement among patches that have been well studied in the context of animal grouping \cite{okubo2001diffusion, turchin1998quantitative, turchin1993quantifying} and are known to facilitate spontaneous organization into heterogeneous patterns \cite{fahimipour2022sharp, anderson2012directional, maimaiti2023stability, lewis1994spatial}. Organisms may also emigrate from habitat patches in groups \cite{cote2017behavioural} and the cues used to synchronize group movements may include those from other species \cite{gil2017social, fahimipour2023wild}. While grouping behavior and synchronized dispersal among group members are likely to also lead to pattern formation, the heterogeneous patterns that result may differ strongly from those in the systems we study here \cite{nakao2010turing, brechtel2018master}.

Teasing apart the contributions of internal and external drivers of observed spatial patterns continues to be a major empirical challenge. Dispersal shaped by interactions with other species is one of the many mechanisms that can generate self-organized variation in community compositions. Given the central role of dispersal behavior in how organisms respond to their environment, future changes to environmental conditions may simultaneously alter external and internal drivers of spatial heterogeneity. Modeling studies such as ours can play an important role in understanding under what conditions internal pattern formation is possible, guiding empirical research.

\subsection*{Data accessibility}
Code to reproduce all data and analyses is publicly available at \cite{lawton2023interspecific}. 

\subsection*{Authors' contributions}
All authors contributed to research design and the drafting, writing and editing of the manuscript. P.L. and A.K.F. implemented data generation and analysis.

\subsection*{Acknowledgements}
A.K.F. was supported by National Science Foundation grant EF-2222478 and K.E.A. was supported by National Science Foundation grant DEB-2225098. We acknowledge this research was primarily conducted on unceded land of the Cahuilla, Tongva, Luiseño, and Serrano peoples native to Southern California. 

\begingroup
\setlength\bibitemsep{0pt}
\printbibliography
\endgroup

\section*{Supplement}
\subsubsection*{Robustness definitions}
First we define a binary variable $\Omega_b(\bm{\phi})$ which equals $1$ if the $n$-dimensional vector of model parameters $\bm{\phi}$ yields a shape of $\lambda(\kappa)$ corresponding to dynamical behavior $b$ and is 0 otherwise. The fraction of parameter space resulting in behavior $b$, i.e. the robustness, is then approximated via the mean value theorem over $N$ random parameter set samples $\bm{\phi}_s$ as 
\setcounter{equation}{0}
\renewcommand\theequation{S\arabic{equation}}
\begin{align}
    \omega_b=\frac{1}{N}\sum_s^N{\Omega_b(\bm{\phi}_s)}.
\end{align}
For the local robustness $\omega_b(\bf{P})$, $\bm{\phi}$ contains the parameters appearing in Eq. 5 for the analysis of 3 species motifs, which are then sampled over $\mathcal{U}(0, 10)$. For the spatial robustness $\omega_b(\bf{C})$, $\bm{\phi}$ contains all potentially nonzero elements of the connectivity matrix $\bf{C}$. As the dynamical behavior is determined by the relative magnitude of linear coefficients appearing in $\bf{C}$ \cite{hata2014sufficient}, we sample elements of $\bf{C}$ on the unit $n-1$ hypersphere, where $n$ is the number of nonzero $\bf{C}$ elements. Finally, for the total robustness $\omega_b(\bf{P},\bf{C})$, $\bm{\phi}$ contains all possible model parameters and the average is taken over all parameter combinations. 

\renewcommand{\thefigure}{S\arabic{figure}}
\setcounter{figure}{0}
\begin{figure*}[t!]
\includegraphics[width=0.95\textwidth]{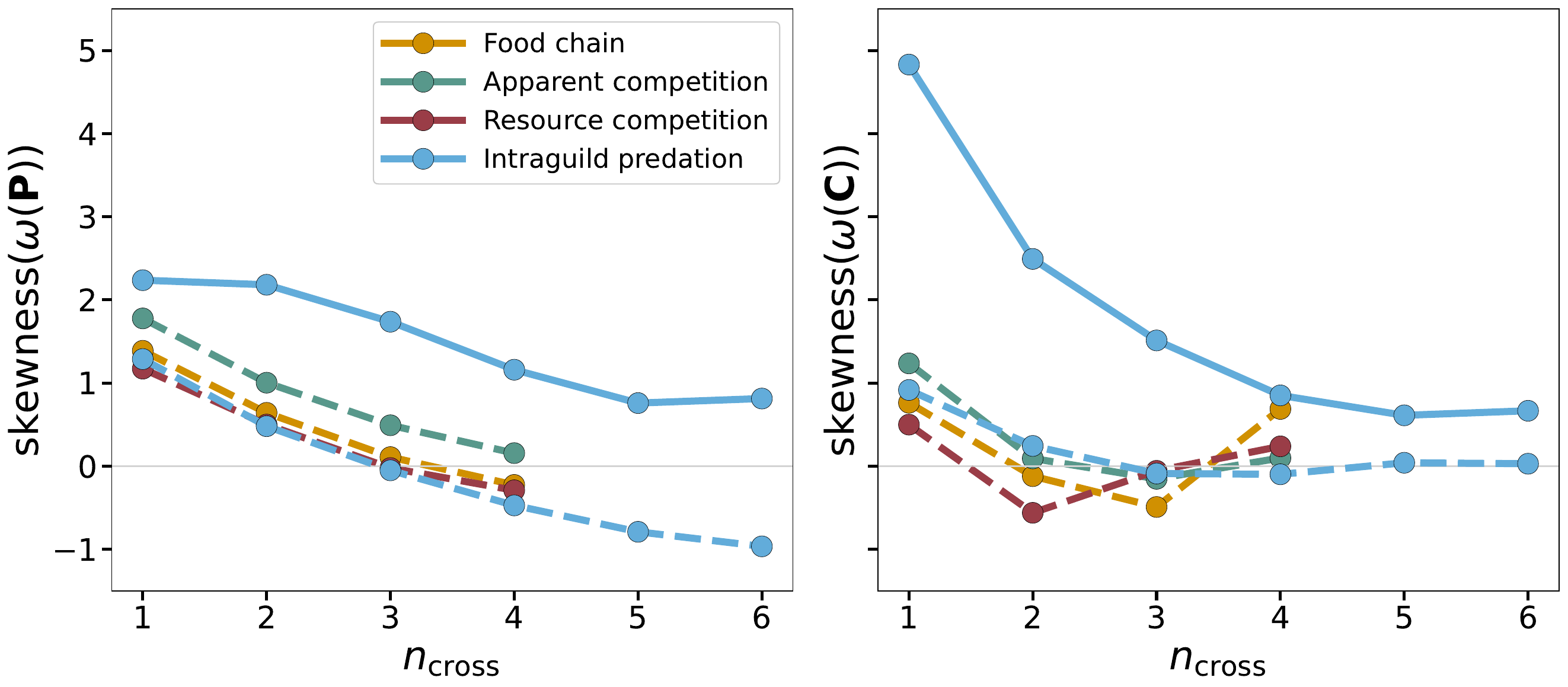}
\caption{Skewness of robustness distributions}{
\footnotesize
\textcolor{mySig}{
Skewness of pattern forming dynamics robustness distributions. Distributions taken over locally feasible and stable webs as done for Fig. 3 in the main text. The constrained skewness is only shown for the intraguild predation motif, as pattern formation is prohibited under cross diffusive constraints in all other motifs, and the skewness is thus uninformative.
}}
\end{figure*}

\end{document}